\documentclass[aps,pra,groupedaddress,showpacs]{revtex4}

\usepackage{graphics}
\usepackage{epsfig}
\usepackage{amsmath}

\begin{document}

\title{Alternative analysis to  perturbation theory}
\author{Juan Mart\'inez-Carranza}
 \affiliation{INAOE, Apartado Postal 51 y 216, 72000, Puebla, Pue., Mexico}
\author{Francisco Soto-Eguibar}
\affiliation{INAOE, Apartado Postal 51 y 216, 72000, Puebla, Pue., Mexico}
\email{feguibar@inaoep.mx}
\author{H\'ector Moya-Cessa}
\affiliation{INAOE, Apartado Postal 51 y 216, 72000, Puebla, Pue., Mexico}
\date{\today}

\begin{abstract}
We develop an alternative approach to  time independent perturbation theory in non-relativistic quantum mechanics. 
The method developed has the advantage to provide in one operation the correction to the energy and to the wave function,
additionally we can analyze the time evolution of the system. To verify our results, we apply our method to the harmonic oscillator perturbed by a 
quadratic potential. An alternative form of the Dyson series, in matrix form instead of integral form, is also obtained.
\end{abstract}

 \maketitle

\section{Introduction}
The Schr\"odinger equation  \cite{1} is the main equation in non-relativistic quantum mechanics. Although it has been widely studied since its introduction,  
there are only few cases in which it can be solved exactly. Typical examples of potentials where an exact analytical solution is known, are the infinite well, 
the harmonic oscillator, the hydrogen atom  \cite{2,3} and the Morse potential  \cite{4}. The great majority of the problems related with the Schr\"odinger 
equation are very complex and can not be solved exactly, as is the case of the cosine potential  \cite{5} for instance. When an exact analytical solution 
can not be found, we are forced to apply approximation methods  \cite{3,6}, that when correctly used, give us a very good understanding of the behavior of the quantum system.
For the sake of clarity, we will briefly revise one of those methods: the time independent perturbation theory, also known as Rayleigh-Schr\"odinger 
perturbation theory. This method has its roots in the works of Rayleigh and Schr\"odinger, but the mathematical foundations were only set by Rellich 
in the late thirties of the past century (see Simon  \cite{7} and references there in). This method has been applied with great success to solve
 a vast variety of problems such that, through its continuous implementation, a lot of techniques have been developed, which go from numerical 
methods  \cite{8} to those more mathematical and fundamental, as  convergency problems  \cite{9,10}.
The Rayleigh-Schr\"odinger theory is appropriate when we have a time independent Hamiltonian, that can be separated in two parts, as follows:
\begin{equation}\label{010}
\hat H=\hat H_0+ \hat H_\textrm{p},
\end{equation}
where $\hat H_0$ is the so-called non-perturbed Hamiltonian, and it is usually assumed to have  known solutions, this is, its eigenvalues
\begin{equation}\label{020}
\hat H_0|n^{(0)}\rangle = E_n^{(0)}|n^{(0)}\rangle
\end{equation}
are known.
The second part of the Hamiltonian, $\hat H_\textrm{p}$, is {\it small} compared to $H_0$; thus $\hat H_\textrm{p}$ is called the "perturbation", because
its effect in the energy spectrum and in the eigenfunctions will be small. To be more  explicit, it is usual to write $\hat H_\textrm{p}$ in terms of a dimensionless  real parameter  
$\lambda$, which is considered very small compared to one
\begin{equation}\label{030}
    \hat H_\textrm{p}=\lambda \hat V \qquad  (\lambda\ll1).
\end{equation}

Here, we propose  an alternative perturbation method based on the evolution operator for Hamiltonian (\ref{010}), i.e., the exponential of the complete Hamiltonian. 
Making use of the Taylor series of the exponential and operator techniques we will cast the solution into a form that has powers of tridiagonal matrices in it. By using 
as an example a perturbative potential for which we know the solutions, we will be able to compare our method with the  Rayleigh-Schr\"odinger perturbation theory. 
Our method will allow also to cast the Dyson series, which are  written in multiple integral forms, as a series of powers of tridiagonal matrices.

We will proceed then as follows, in section II, we present a brief summary of the standard time independent perturbation theory, emphasizing the expressions obtained 
for the first and second order corrections. As is well known, two expressions are obtained at each order, one for the energies and one for the wave functions. In Section 
III, we introduce our matrix approach to perturbation theory that will allow us to obtain a single solution that contains both, the energy and wave function corrections. 
We will introduce there all the corrections to the wave function that will allow us to generate later the Dyson series, and in this way give also a new expression of it but now 
in terms of a matrix series.  In Section IV we compare the our method with the standard perturbation theory by solving the harmonic oscillator with a quadratic perturbation. 
Although this  seems redundant, because this case  has an exact solution, it will allow us to compare both methods by doing an expansion of the exact solution in terms of the 
perturbation parameter, $\lambda$. In Section V we apply our formalism to rewrite the Dyson series in matrix form and Section VI is left for conclusions.

\section{Perturbation theory}
As we already mention, we have to solve the eigenvalue problem given by (\ref{010}). Standard perturbation theory produces the following expressions for the eigenvalues
\begin{equation}\label{050}
    E_n=E_n^{(0)}+\lambda \Delta_n^{(1)}+\lambda^2 \Delta_n^{(2)}+\cdots,
\end{equation}
and  the eigenfunctions 
\begin{equation}\label{060}
    |n\rangle=|n^{(0)}\rangle+\lambda |n^{(1)}\rangle+\lambda^2 |n^{(2)}\rangle+\cdots.
\end{equation}
In the two previous equations the super index $(j)$ indicates the correction order.
The first and second order corrections for the energy are
\begin{eqnarray}\label{080}
 \Delta_n^{(1)}&=&\langle n^{(0)}|\hat V|n^{(0)}\rangle=V_{nn}\label{070}\\
 \Delta_n^{(2)}&=&\langle n^{(0)}|\hat V|n^{(1)}\rangle=\sum_{k\neq n}\frac{|V_{kn}|^2}{E_n^{(0)}-E_k^{(0)}}
\end{eqnarray}
where we have defined
\begin{equation}\label{090}
    V_{kn}=\langle k^{(0)}|\hat V|n^{(0)}\rangle.
\end{equation}
For the wave functions the first two order corrections are given by the expressions
\begin{equation}\label{100}
   |n^{(1)}\rangle = \sum_{k\neq n}\frac{|k^{(0)}\rangle V_{kn}}{E_n^{(0)}-E_k^{(0)}}
\end{equation}
and
\begin{equation}\label{110}
 |n^{(2)}\rangle = \sum_{k\neq n}\sum_{m\neq n}\frac{|{k^{(0)}}\rangle V_{km}V_{mn}}{(E_n^{(0)} -E_k^{(0)})(E_n^{(0)} -E_m^{(0)})}
- \sum_{k\neq n}\frac{|{k^{(0)}}\rangle V_{nn}V_{kn}}{(E_n^{(0)}
-E_k^{(0)})^2}
\end{equation}
It is clear that the previous expressions can be used in the non-degenerate case, where we always have $E_n^{(0)}\neq E_k^{(0)}$. In the degenerate case, a different treatment is needed.

\section{Matrix approach to the perturbation theory}
\subsection{First order correction}
We now find an approximate solution for the complete (time dependent) Schr\"odinger equation; i.e., we will solve approximately the equation
\begin{equation}\label{120}
i\frac{\partial }{\partial t}|\psi\rangle=\hat H|\psi\rangle=\left(\hat H_0 +\lambda\hat V\right)|\psi\rangle.
\end{equation}
The formal solution may be written as
\begin{equation}\label{130}
|\psi(t)\rangle=e^{-it(\hat H_0+\lambda\hat V)}|\psi(0)\rangle ,
\end{equation}
that by expanding the exponential in a Taylor series
\begin{equation}\label{140}
|\psi(t)\rangle= \sum_{n=0}^{\infty}\frac{(-it)^n}{n!}\left(
\hat H_0  +\lambda\hat V \right)^n |\psi(0)\rangle,
\end{equation}
is obtained. 
If we develop the binomials inside the summation, rearrange  terms and  cut the series to first order in $\lambda$, we have
\begin{equation}\label{150}
|\psi(t)\rangle \approx  \left(\sum_{n=0}^{\infty}\frac{(-it)^n}{n!}\hat H_0^n+
\lambda\sum_{n=1}^{\infty}\frac{(-it)^n}{n!}\sum_{k=0}^{n-1} \hat
H_0^{n-1-k}\hat V \hat H_0^k \right) |\psi(0)\rangle.
\end{equation}
The key ingredient of  the method we introduce in this contribution is the matrix
\begin{equation}\label{160}
M= \left( \begin{array}{cccccc}
\hat{H}_0 & \hat V   \\
0         & \hat{H}_0 \\
\end{array} \right).
\end{equation}
It is very easy to get convinced that the following relations are satisfied
\begin{eqnarray}
(M)_{1,2}&=&\hat V ,\nonumber\\
(M^2)_{1,2}&=&\hat H_0\hat V+\hat V\hat H_0,\nonumber\\
&\vdots &\nonumber\\
(M^n)_{1,2}&=&\sum_{k=0}^{n-1} \hat H_0^{n-1-k}\hat V \hat
H_0^k\nonumber.
\end{eqnarray}
Thus, equation (\ref{150}) can be written as
\begin{equation}\label{170}
|\psi(t)\rangle \approx \left[\sum_{n=0}^{\infty}\frac{(-it)^n}{n!}(\hat{H}_0)^n
+\lambda\sum_{n=1}^{\infty}\frac{(-it)^n(M^n)_{(1,2)}}{n!}\right]|\psi(0)\rangle.
\end{equation}
As $M^0=I$ (where $I$ is the [$2\times2$]  identity matrix), we trivially have that
\begin{equation*}
(M^0)_{(1,2)}=0,
\end{equation*}
and substituting this term in equation (\ref{170}) we obtain
\begin{equation}\label{180}
|\psi(t)\rangle\approx \left(e^{-i\hat{H}_0 t} +\lambda \left(
e^{-iMt} \right) _{(1,{2})}\right)|\psi(0)\rangle
=e^{-i\hat{H}_0 t} |\psi(0)\rangle
 +\lambda \left(e^{-iMt} \right) _{(1,{2})}|\psi(0)\rangle.
\end{equation}
Note that in the last expression we have separated the approximate solution in two parts; the first part is the solution of the non-perturbed system, that is well 
known, and the second part is the first order correction to the wave function. We now show how the correction to first order may be calculated, for this we rewrite equation (\ref{180}) as
\begin{equation}\label{190}
|\psi(t)\rangle\approx|\psi^{(0)}(t)\rangle+\lambda( ||\psi^{p}\rangle)_{1,2},
\end{equation}
where we have defined the {\it matrix wave  function}
\begin{equation}\label{200}
||\psi^{p}\rangle= \left( \begin{array}{cccccc}
|\psi_{(1,1)}\rangle & |\psi_{(1,2)}\rangle \\
|\psi_{(2,1)}\rangle & |\psi_{(2,2)}\rangle \\
\end{array} \right).
\end{equation}

Deriving (\ref{180}) and (\ref{190}) with respect to time and equating them, we arrive to the equation
\begin{equation}\label{210}
i \frac{\partial }{\partial t}|\psi^{(0)}(t)\rangle+i\lambda
\frac{\partial }{\partial
t}(||\psi^{p}\rangle)_{1,2}=\hat{H}_0e^{-i\hat{H}_0
t}|\psi(0)\rangle +\lambda \left( Me^{-it M}
I|\psi(0)\rangle\right)_{1,2}.
\end{equation}
 Equating powers of $\lambda$, we have
\begin{equation}\label{220}
i \frac{\partial }{\partial t}\left(
|\psi^{p}\rangle\right)_{1,2}=\left( Me^{-it M}
I|\psi(0)\rangle\right)_{1,2} .
\end{equation}
We have to solve now equation (\ref{220}), or equivalently the system
\begin{equation}
 i\frac{\partial}{\partial t}\left( \begin{array}{cccccc}
|\psi_{(1,1)}\rangle & |\psi_{(1,2)}\rangle  \\
|\psi_{(2,1)}\rangle & |\psi_{(2,2)}\rangle  \\
\end{array} \right) = Me^{-itM}\left( \begin{array}{cccccc}
|\psi(0)\rangle & 0 \\
0               & |\psi(0)\rangle \\
\end{array} \right).
\end{equation}
Integrating this equation, we have
\begin{equation}\label{240}
||\psi^{p}\rangle =e^{-it M}\left( \begin{array}{ccc}
|\psi(0) \rangle & 0 \\
0 & |\psi(0) \rangle \\
 \end{array} \right).
\end{equation}
to finally write the differential equation 
\begin{equation}\label{250}
i\frac{\partial }{\partial t} ||\psi^{p}\rangle=M ||\psi^{p}\rangle,
\end{equation}
with the initial condition
\begin{equation}\label{260}
||\psi^{p}(0)\rangle=\left( \begin{array}{cccccc}
|\psi(0)\rangle & 0                \\
0               & |\psi(0)\rangle  \\
\end{array} \right) .
\end{equation}
The needed solution is associated with the second column of matrix $||\psi^p \rangle$, thus we write
\begin{equation*}
 i\frac{\partial}{\partial t}\left( \begin{array}{cccccc}
 |\psi_{(1,2)}\rangle  \\
 |\psi_{(2,2)}\rangle  \\
\end{array} \right) = M \left( \begin{array}{cccccc}
 |\psi_{(1,2)}\rangle  \\
 |\psi_{(2,2)}\rangle  \\
\end{array} \right),
\end{equation*}
but, because $M$ is a tridiagonal matrix, ithe system may be directly integrated. We show it by writing explicitly  
\begin{equation}\label{270}
  i\frac{\partial}{\partial t}|\psi_{(1,2)}\rangle = \hat H_0|\psi_{(1,2)}\rangle + \hat V|\psi_{(2,2)}\rangle ,\
\end{equation}
and
\begin{equation}\label{280}
   i\frac{\partial}{\partial t}|\psi_{(2,2)}\rangle = \hat H_0|\psi_{(2,2)}\rangle,\
\end{equation}
with the initial condition
\begin{equation}\label{290}
\left( \begin{array}{cccccc}
 0  \\
 |\psi (0)\rangle  \\
\end{array} \right).
\end{equation}
Because we know the solution for $\hat{H}_0$, equation (\ref{280}) is solved trivially,
\begin{equation}\label{300}
|\psi_{(2,2)}\rangle = e^{-i t \hat H_0}|\psi (0)\rangle,
\end{equation}
that substituted in (\ref{270}), allows us to write
\begin{equation*}
i\frac{\partial}{\partial t}|\psi_{(1,2)}\rangle =\hat H_0|\psi_{(1,2)}\rangle + \hat Ve^{-i\hat H_0 t}|\psi (0)\rangle .
\end{equation*}
Making the transformation $|\psi_{(1,2)}\rangle = e^{-i\hat H_0}|\phi_{1,2}(x)\rangle$, we arrive to the equation
\begin{equation*}
  i\frac{\partial }{\partial t}|\phi_{1,2}\rangle=e^{i\hat H_0 t}\hat V^{-i\hat H_0 t}|\psi(0)\rangle\,
\end{equation*}
that can be easily  integrated to give
\begin{equation*}
 |\phi_{1,2}\rangle=-i\left[\int_0^{t} e^{i\hat H_0 t_1}\hat Ve^{-i\hat H_0 t_1}  dt_1 \right]|\psi(0)\rangle,
\end{equation*}
and by transforming back  the solution to the first correction may be obtained
\begin{equation} \label{310}
| \psi_{1,2}\rangle=-ie^{-i\hat{H}_0 t}\left[\int_0^{t} e^{i\hat{H}_0 t_1} \hat V e^{-i\hat{H}_0 t_1}dt_1 \right] |\psi(0)\rangle.
\end{equation}

Up to here we have produced a first order correction for the wave function with no assumptions on Hamiltonian degeneracy, therefore making this first order correction valid 
also for degenerate Hamiltonians.

As the eigenfunctions of the non-perturbed Hamiltonian constitutes a complete orthonormal set, we can write the closure condition as $
\hat I =\sum_{k}|k^{(0)}\rangle\langle k^{(0)}|
$, where $|k^{(0)}\rangle$ are the eigenstates of the unperturbed Hamiltonian
and insert the identity operator written in this way, in equation (\ref{310}) to arrive to
\begin{equation}\label{330}
  | \psi_{1,2}\rangle = -i \sum_{k} |k^{(0)}\rangle \langle k^{(0)}| e^{-i\hat{H}_0 t}
  \left[\int_0^{t} e^{i\hat{H}_0 t_1} \hat V e^{-i\hat{H}_0 t_1} dt_1 \right] |\psi (0)\rangle.
\end{equation}
For simplicity we set the initial condition $|\psi(0)\rangle=|n^{(0)}\rangle$.  By defining
\begin{equation}\label{037}
    |n^{(1)}\rangle_t=\sum_{k\neq n}\frac{|k^{(0)}\rangle V_{kn}}{E^{(0)}_n- E^{(0)}_k}e^{-iE_k^{(0)} t},
\end{equation}
and using expressions (\ref{070}) and (\ref{100}), we finally get
\begin{eqnarray}\label{380}
| \psi_{1,2}\rangle &=&e^{-iE_n^{(0)} t}\left[|n^{(1)}\rangle-it|n^{(0)}\rangle
\Delta_n^{(1)}\right]-|n^{(1)}\rangle_t.
\end{eqnarray}
We  note that the first order corrections to the energy and to the wave function are already contained in the last expression. Additionally we can show that we can write 
\begin{equation}\label{39}
   | |\psi^{p}\rangle= e^{-it\hat H_0}\left( \begin{array}{cccccc}
   1 & -i\int_0^{t}  dt_1\hat V(t_1)   \\
     0               & 1                 \\
  \end{array} \right)|\psi(0)\rangle,
\end{equation}
with
\begin{equation}
\hat V(t_1)=e^{i\hat H_0 t_1}\hat V e^{-i\hat H_0 t_1}.
\end{equation}
\subsection{Second order correction}
We will find now a similar expression for the second order correction. Expanding again (\ref{140}) in Taylor series  and keeping terms to $\lambda^2$, we have
\begin{eqnarray}\label{400}
\nonumber    |\psi(t)\rangle &\approx& \left( \sum_{n=0}^{\infty} \frac{(-it)^n}{n!}\hat H_0^n+ \lambda\sum_{n=1}^{\infty}\frac{(-it)^n}{n!}\sum_{k=0}^{n-1} \hat H_0^{n-1-k}\hat V \hat H_0^k \right)|\psi(0)\rangle \\
     &+& \lambda^2\sum_{n=2}^{\infty}\frac{(-it)^n}{n!}\sum_{k=1}^{n-1}\sum_{j=0}^{n-k}\hat
    H_0^{n-k-j-1}\hat V\hat H_0^{j}\hat V \hat H_0^{k-1}
|\psi(0)\rangle .
\end{eqnarray}
In analogy with the first order correction, we define now the matrix
\begin{equation}\label{410}
    M= \left( \begin{array}{cccccc}
            \hat{H}_0 & \hat V     &  0       \\
                     0         & \hat{H}_0  &  \hat V   \\
               0         &       0    &  \hat{H}_0\\
               \end{array} \right).
\end{equation}
It is very easy to see that in this case 
\begin{eqnarray*}
(M^2)_{1,3}&=&\hat V^2 , \\
(M^3)_{1,3}&=&\hat H_0\hat V^2+\hat V\hat H_0\hat V+\hat V^2\hat H_0, \\
&\vdots &  \\
(M^n)_{1,3}&=&\sum_{k=1}^{n-1}\sum_{j=0}^{n-k}\hat
H_0^{n-k-j-1}\hat V\hat H_0^{j}\hat V(\hat H_0)^{k-1};
\end{eqnarray*}
so we can write equation (\ref{400}) as
\begin{equation}\label{420}
|\psi(t)\rangle \approx   \left(e^{-i\hat{H}_0 t} + \lambda \left(
e^{-iMt} \right)
_{(1,{2})}+\lambda^2\sum_{n=2}^{\infty}\frac{(-it)^n}{n!}\left(M^n
\right)_{1,3}\right)|\psi(0)\rangle,
 \end{equation}
and knowing that
\begin{equation*}
    \left(M^0 \right)_{1,3}=\left(M \right)_{1,3}=0,
\end{equation*}
we obtain
\begin{eqnarray}\label{430}
 |\psi(t)\rangle&\approx & \left(e^{-i\hat{H}_0 t} +\lambda \left( e^{-iMt} \right) _{(1,{2})}+\lambda^2\left( e^{-iMt} \right) _{(1,{3})}\right)|\psi(0)\rangle.
 \end{eqnarray}
Inserting now the matrix
\begin{equation}\label{440}
||\psi^{p}\rangle= \left( \begin{array}{cccccc}
|\psi_{(1,1)}\rangle & |\psi_{(1,2)}\rangle & |\psi_{(1,3)}\rangle \\
|\psi_{(2,1)}\rangle & |\psi_{(2,2)}\rangle & |\psi_{(2,3)}\rangle \\
|\psi_{(3,1)}\rangle & |\psi_{(3,2)}\rangle & |\psi_{(3,3)}\rangle \\
\end{array} \right).
\end{equation}
where the first order corrections $|\psi_{(1,2)}\rangle$ and the second order corrections $|\psi_{(1,3)}\rangle$  are included.\\
Expanding $|\psi\rangle$ to second order in $\lambda$, we get
\begin{equation}\label{450}
|\psi\rangle  \approx |\psi^{(0)}(t)\rangle+\lambda(||\psi^{p}\rangle)_{1,2}+\lambda^2( || \psi^{p}\rangle)_{1,3}.
\end{equation}
Following the same procedure as in the  first order correction case, we derive equations (\ref{430}) and (\ref{450}) with respect to time, and equate the corresponding equations to obtain
\begin{eqnarray}\label{460}
\nonumber
i \frac{\partial }{\partial t}|\psi^{(0)}(t)\rangle+i\lambda
\frac{\partial }{\partial t}(||\psi^{p}\rangle)_{1,2}+\lambda^2
\frac{\partial }{\partial
t}(||\psi^{p})\rangle_{1,3}&=&\hat{H}_0e^{-i\hat{H}_0
t}|\psi(0)\rangle +\lambda \left( Me^{-it M}
I|\psi(0)\rangle\right)_{1,2} \\ &+&\lambda^2 \left( Me^{-it M}
I|\psi(0)\rangle\right)_{1,3}.
\end{eqnarray}
Equating powers of $\lambda^2$, we can establish that
\begin{equation*}
i\frac{\partial}{\partial t}(||\psi^p\rangle)_{1,3}=\left(Me^{-itM}I|\psi(0)\rangle\right)_{1,3}.
\end{equation*}
or equivalently
\begin{equation}\label{470}
i\frac{\partial}{\partial t}||\psi^p\rangle=M||\psi^p\rangle,
\end{equation}
with the initial condition
\begin{equation}\label{480}
||\psi^p (0)\rangle=\left( \begin{array}{ccc}
|\psi(0)\rangle & 0 & 0 \\
0 & |\psi(0)\rangle & 0 \\
0 & 0 & |\psi(0)\rangle \\
\end{array}\right).
\end{equation}
Equation (\ref{480}) is similar to (\ref{250}), thus we can again proceed as in the first order case, choosing the third column in both sides of the equation and getting the differential equations system
\begin{equation}\label{490}
 i\frac{\partial}{\partial t}\left( \begin{array}{cccccc}
|\psi_{(1,3)}\rangle  \\
|\psi_{(2,3)}\rangle\\
|\psi_{(3,3)}\rangle
\end{array} \right) = M\left( \begin{array}{cccccc}
 |\psi_{(1,3)}\rangle  \\
 |\psi_{(2,3)}\rangle  \\
 |\psi_{(3,3)}\rangle
\end{array} \right),
\end{equation}
with the initial condition
\begin{equation}\label{500}
\left( \begin{array}{cccccc}
 0  \\
 0  \\
 |\psi (0)\rangle
\end{array} \right).
\end{equation}
The correction we were looking for is then
\begin{equation}\label{510}
 |\psi_{1,3}\rangle=-ie^{-i\hat H_0 t}\int_{0 }^{t} e^{i\hat H_0 t_1}\hat V\left[-ie^{-i\hat H_0t_1}\int_{0}^{t_1} e^{i\hat H_0 t_2}\hat Ve^{-i\hat H_0 t_2}dt_2|\psi(0)\rangle\right] dt_1.
\end{equation}
In this equation we note that the expression inside the square brackets is the first order correction, then using (\ref{380}) we can write
 \begin{equation}\label{520}
 |\psi_{1,3}\rangle=-ie^{-i\hat H_0 t}\int_{0 }^{t} e^{i\hat H_0 t_1}\hat V\left[e^{-it_1 E_{n}^{(0)}}
 \left(|n^{(1)}\rangle-it_1|n^{(0)}\rangle\Delta_n^{(1)}-|n^{(1)}\rangle_{t_1}\right)\right]dt_1.
 \end{equation}
Inserting the identity operator $\hat I$ as we had defined it in the first order correction, we have
\begin{eqnarray}\label{530}
     |\psi_{1,3}\rangle&=&-i\sum_{k}|k^{(0)}\rangle\langle k^{(0)}|e^{-i\hat H_0 t}\int_{0 }^{t} e^{i\hat H_0 t_1}\hat V\left[e^{-it_1\omega_{n}^{(0)}}\left(|n^{(1)}\rangle-it_1|n^{(0)}\rangle\Delta_n^{(1)}\right)-|n^{(1)}\rangle_{t_1}\right]dt_1,\nonumber\\
     &=&e^{-i\hat E^{(0)}_n t}\Bigg[\sum_{k\neq n}\sum_{m\neq n}\frac{|{k^{(0)}}\rangle V_{km}V_{mn}}{(E_n^{(0)} -E_k^{(0)})(E_n^{(0)} -E_m^{(0)})}
    - \sum_{k\neq n}\frac{|{k^{(0)}}\rangle \Delta^{(1)}_n V_{kn}}{(E_n^{(0)} -E_k^{(0)})^2}-it\langle n^{(0)}|\hat V |n^{(1)}\rangle\nonumber\\
&+& it^2(\Delta^{(1)}_n)^2 -\sum_{k\neq n}\frac{it|{k^{(0)}}\rangle \Delta^{(1)}_n V_{kn}}{E_n^{(0)} -E_k^{(0)}}\Bigg]-|{n^{(2)}}\rangle_t+|{n^{(2)}_{1}}\rangle_t,
 \end{eqnarray}
with
\begin{equation}\label{540}
    |{n^{(2)}}\rangle_t =\sum_{k\neq n}\sum_{m\neq n}\Bigg[\frac{|{k^{(0)}}\rangle V_{km}V_{mn}}{(E_n^{(0)} -E_k^{(0)})(E_n^{(0)} -E_m^{(0)})}
  - \sum_{k\neq n}\frac{|{k^{(0)}}\rangle \Delta^{(1)}_n
    V_{kn}}{(E_n^{(0)} -E_k^{(0)})^2}-\sum_{k\neq
   n}\frac{it|{k^{(0)}}\rangle \Delta^{(1)}_n V_{kn}}{E_o n^{(0)}
   -E_k^{(0)}}\Bigg]e^{-i\hat E^{(0)}_k t}
\end{equation}
and
\begin{eqnarray}
|{n^{(2)}_{1}}\rangle_t &= &i\sum_{k}|k^{(0)}\rangle\langle k^{(0)}|e^{-i\hat H_0 t}\int_{0 }^{t} e^{i\hat H_0 t_1} \hat V|n^{(1)}(t_1)\rangle dt_1,\nonumber\\
&=&\sum_{k}\sum_{m\neq n}\frac{|k^{(0)}\rangle e^{-itE_k^{(0)}} V_{km}V_{mn}}{E_k^{(0)} -E_m^{(0)}}\int_{0}^{t}e^{it(E^{(0)}_k-E^{(0)}_m)}dt_1\label{541}
\end{eqnarray}
In the same way as in the first order correction case, equation (\ref{530}) gives us the correction to the wave function and to the energy in one operation; 
however, it generates the additional expressions $|{n^{(2)}}(t)\rangle$ y $|{n^{(2)}_{1}}(t)\rangle$. \\
We can show that it may be written in the compact form
\begin{equation}\label{550}
||\psi^{p}\rangle= e^{-it\hat H_0}\left( \begin{array}{cccccc}
1 & -i\int_0^{t}  dt_1\hat V(t_1) &(-i)^2\int_0^{t}dt_1\int_0^{t_1} dt_2 \hat V(t_1)\hat V(t_2)  \\
0               & 1       & -i\int_0^{t}  dt_1\hat V(t_1)           \\
0               & 0         & 1
\end{array} \right)|\psi(0)\rangle.
\end{equation}

\subsection{Higher order corrections}
In the previous sections, we have found the corrections to first and second order, and we have shown that the results are not only consistent with the traditional 
perturbation theory, but new terms arise. These new terms will be shown correct in Section IV where we compare both solutions.
In rest of this section we generalize our method to higher orders. We propose the perturbation series
\begin{eqnarray}\label{560}
|\psi^{(0)}(t)\rangle +
\sum_{n=1}^{m}\lambda^n(||\psi^{p}\rangle)_{1,m+1} &\approx &
\left(e^{-it\hat
H_0}+\sum_{n=1}^{m}\lambda^n\left(e^{-itM}\right)\right)|\psi(0)\rangle
\end{eqnarray}
with
\begin{equation}\label{570}
||\psi^{p}\rangle=\left(
\begin{array}{cccc}
|\psi_{(1,1)}\rangle & \cdots & |\psi_{(1,m+1)}\rangle\\
\vdots               & \ddots & \vdots                \\
|\psi_{(m+1,1)}\rangle & \cdots & |\psi_{(m+1,m+1)}\rangle
\end{array}\right),
\end{equation}
where $m$ is the correction order, and
\begin{equation}\label{580}
M=\left(
\begin{array}{cccc}
\hat H_0 & \hat V & \cdots & 0               \\
0 & \hat H_0 & \cdots & 0               \\
\vdots          &        & \ddots & \vdots    \\
0               & \cdots &        & \hat H_0
\end{array}\right).
\end{equation}
Following the method for the first and second order corrections, we deduce the following system of differential equations
\begin{equation}\label{590}
  i\frac{\partial}{\partial t}(||\psi^{p}\rangle)_{1,m+1}=M(||\psi^{p}\rangle)_{1,m+1},
\end{equation}
or
\begin{equation}
i\frac{\partial }{\partial t}\left( \begin{array}{ccccc}
|\psi_{1,m+1}\rangle\\
\vdots            \\
|\psi_{m+1,m+1}\rangle
 \end{array} \right)=M
 \left( \begin{array}{ccccc}
|\psi_{1,m+1}\rangle\\
\vdots            \\
|\psi_{m+1,m+1}\rangle
 \end{array} \right)
,\label{55}
\end{equation}
with the initial condition
\begin{equation}\label{600}
 \left( \begin{array}{ccccc}
0\\
\vdots            \\
|\psi(0)\rangle
 \end{array} \right),
\end{equation}
so the solution we are looking for is
\begin{equation}\label{620}
 |\psi_{1,n+1}\rangle=-ie^{-iH_0 t}\int_{0 }^{t} e^{iH_0
t_1}V\left[-ie^{-iH_0t_1}\int_{0}^{t_1} e^{iH_0
t_2}V\left[-ie^{-iH_0 t_2}\int_0^{t_2}\cdots dt_3\right]
dt_2|\psi(0)\rangle\right] dt_1.
\end{equation}

\section{Comparison of the standard and the matrix methods}
We will treat now the case of a harmonic oscillator perturbed by a quadratic potential. More than as an example, what we pretend in this section 
is to show that in this case, where an exact analytic solution is known, our method gives a {\it better }solution than the standard Rayleigh-Schr\"odinger perturbation theory.
The non-perturbed Hamiltonian is
\begin{equation}\label{630}
    \hat H_0=\frac{1}{2}(\hat p^2 + \omega^2\hat {x}^2),
\end{equation}
and the perturbation potential is
\begin{equation}\label{640}
    \hat V=\frac{\omega^2}{2}\hat x^2;
\end{equation}
for simplicity we consider a unity mass oscillator\\
The total Hamiltonian is
\begin{equation}\label{650}
\hat H_{\tilde\omega}=\frac{1}{2}(\hat p^2 +\tilde\omega^2 \hat x^2),
\end{equation}
with
\begin{equation}\label{655}
\tilde\omega=\omega\sqrt{1+\lambda}.
\end{equation}
Physically this Hamiltonian can represent a one mode cavity, to which the oscillation frequency can be changed by changing the perturbation parameter. The final result of a modification 
in the initial frequency produces squeezed coherent states   \cite{moya}, as it is described by Dutra   \cite{11}.\\

First we obtain the first order correction with the matrix method we have just introduced. We do so by
substituting the perturbation potential in equation (\ref{310}) and obtain
\begin{equation}\label{660}
   |\psi_{1,2}\rangle=-i\frac{\omega^2}{2}
   e^{-i\hat{H}_0 t} \left[\int_0^{t} e^{i\hat{H}_0 t_1} \hat x ^2 e^{-i\hat{H}_0
   t_1}dt_1 \right] |\psi(0)\rangle.
\end{equation}
By defining the position operator in terms of annihilation and creation operators,  $\hat a$ and $\hat a ^{\dagger}$
\begin{equation}\label{670}
\hat x = \frac{\hat a + \hat a^{\dagger}}{\sqrt{2\omega}},
\end{equation}
we can write equation  (\ref{660}) as
\begin{equation}\label{680}
|\psi_{1,2}\rangle=-i\frac{\omega}{4}e^{-i\hat{H}_0
t}\left[\int_0^{t} e^{i\hat{H}_0 t_1}\left(\hat a + \hat
a^{\dagger}\right)^ 2 e^{-i\hat{H}_0 t_1}dt_1 \right]|\psi(0)\rangle.
\end{equation}
Using the relations
\begin{equation}\label{690}
   e^{i\hat H_0 t}\hat a e^{-i\hat H_0 t}=\hat a e^{-i\omega t},
\end{equation}
and
\begin{equation}\label{700}
e^{i\hat H_0 t}\hat a^{\dagger} e^{-i\hat H_0 t}=\hat a^{\dagger} e^{i\omega t},
\end{equation}
equation (\ref{680}) is transformed in
\begin{eqnarray}\label{710}
|\psi_{1,2}\rangle&=&-i\frac{\omega}{4}e^{-i\hat{H}_0 t}\left[\int_0^{t}\left(\hat a^2 e^{-2i\omega t} +\hat a\hat a^{\dagger}+\hat a^{\dagger}\hat a + (\hat a^{\dagger})^2 e^{2i\omega t}\right) dt_1 \right] |\psi(0)\rangle,\nonumber\\
&=&-i\frac{\omega}{4}e^{-i\hat{H}_0 t}\left[\frac{\hat
a^2}{-2i\omega} \left(e^{-2i\omega t} -1\right)+\frac{(\hat
a^{\dagger})^2}{2i\omega} \left(e^{2i\omega t} -1\right)+t(\hat
a\hat a^{\dagger}+\hat a^{\dagger}\hat a )\right] |\psi(0)\rangle.
\end{eqnarray}
By writing the unperturbed Hamiltonian in terms of the number operator $\hat{n}=\hat a ^{\dagger}\hat a $
and  using $|\psi(0)\rangle= |n^{(0)}\rangle=$ we obtain
\begin{eqnarray}\label{720}
|\psi_{1,2}\rangle &=& e^{-iE_n^{(0)} t}\left[ \left
(\frac{\sqrt{n(n-1)}}{ 8}|n^{(0)}-2^{(0)}\rangle  -
\frac{\sqrt{(n+1)(n+2)}}{8 }|n^{(0)}+2^{(0)}\rangle \right )
-|n^{(0)}\rangle\frac{i\omega}{4}( 2 n +1)t\right]\nonumber\\
&-&\left (e^{ -iE^{(0)}_{n-2} t} \frac{\sqrt{n(n-1)} }{
8}|n^{(0)}-2^{(0)}\rangle - e^{-iE^{(0)}_{n+2} t
}\frac{\sqrt{(n+1)(n+2)} }{8}|n^{(0)}+2^{(0)}\rangle\right ).
\end{eqnarray}
Comparing with (\ref{380}), we have that
\begin{equation}\label{730}
|n^{(1)}\rangle = \frac{\sqrt{n(n-1)}}{ 8}|n^{(0)}-2^{(0)}\rangle  - \frac{\sqrt{(n+1)(n+2)}}{8 }|n^{(0)}+2^{(0)}\rangle,
\end{equation}
\begin{equation}\label{740}
\Delta^{(1)}_n= \frac{\omega}{4}(2n+1),
\end{equation}
and
\begin{equation}\label{750}
|n^{(1)}\rangle_t = e^{ -iE^{(0)}_{n-2} t}
\frac{\sqrt{n(n-1)} }{ 8}|n^{(0)}-2^{(0)}\rangle -
e^{-iE^{(0)}_{n+2} t }\frac{\sqrt{(n+1)(n+2)}
}{8}|n^{(0)}+2^{(0)}\rangle,
\end{equation}
with $E_n^{(0)}=\omega(n+1/2)$. \\
As we can see (\ref{730}) and (\ref{740}) are the corresponding expressions to the corrections to the wave function and to the energy to first order that are 
obtained when the stationary Schr\"odinger equation is considered, while (\ref{750}) is similar to (\ref{730}), but with the difference that this expression has time dependent factors. \\
\subsection{Exact solution}
We may see already differences between both methods. In order to compare them, we need to use an example that may be solved, such that, once we have the exact solution, this can
be expanded on a (perturbation) parameter and see if it matches the results we obtained.

In this case, the exact formal solution is given by
\begin{equation}\label{910}
|\psi(t)\rangle=e^{-{i\hat H_{\tilde{\omega}}}t}|\psi(0)\rangle.
\end{equation}
where $\hat H_{\tilde{\omega}}$ is the complete Hamiltonian defined in (\ref{650}).\\
As usual, we introduce the ladder operators
\begin{eqnarray}\label{940}
\hat A^{\dagger}&=&\sqrt{\frac{\tilde{\omega}}{2}}\hat x-i\frac{\hat p}{\sqrt{2\tilde{\omega}}}
\end{eqnarray}
and
\begin{equation}\label{950}
\hat A=\sqrt{\frac{\tilde{\omega}}{2}}\hat x+i\frac{\hat p}{\sqrt{2\tilde{\omega}}}.
\end{equation}
in terms of which the perturbed Hamiltonian (\ref{650}) can be written as
\begin{equation}\label{970}
\hat H_{\tilde{\omega}}=\tilde{\omega}\left(\hat A^{\dagger}\hat A +\frac{1}{2}\right).
\end{equation}
It is easy to show that
\begin{equation}\label{980}
\hat A^{\dagger}=\sqrt{\frac{\tilde{\omega}}{2}}\frac{\hat a+\hat a^{\dagger}}{\sqrt{2\omega}}-\sqrt{\frac{\omega}{2}}\frac{(\hat a-\hat a^{\dagger})}{\sqrt{2\tilde{\omega}}}
\end{equation}
and
\begin{equation}\label{990}
\hat A=\sqrt{\frac{\tilde{\omega}}{2}}\frac{\hat a+\hat a^{\dagger}}{\sqrt{2\omega}}+\sqrt{\frac{\omega}{2}}\frac{(\hat a-\hat a^{\dagger})}{\sqrt{2\tilde{\omega}}},
\end{equation}
Defining the squeeze operator   \cite{moya}
\begin{equation}\label{1000}
\hat S(\lambda)=\exp \left\{\frac{\ln({1+\lambda})}{8}\left[\hat a^2-(\hat a^{\dagger})^2\right]\right\},
\end{equation}
it can be shown that
\begin{eqnarray}\label{1110}
\hat A&=&\hat S\hat a\hat S^{\dagger}
\end{eqnarray}
and
\begin{eqnarray}\label{1120}
\hat A^{\dagger}&=&\hat S\hat a^{\dagger}\hat S^{\dagger}.
\end{eqnarray}
Thus, the formal solution (\ref{910}) can be written as
\begin{eqnarray}\label{1130}
| \psi(t)\rangle &=&e^{-it \tilde{\omega}(\hat A^{\dagger}\hat A+1/2)}| \psi(0)\rangle,\nonumber\\
 &=&e^{-it \tilde{\omega}\hat S(\hat a^{\dagger}\hat a+1/2)S^{\dagger}}| \psi(0)\rangle,\nonumber\\
 &=&\hat Se^{-it \tilde{\omega}(\hat a^{\dagger}\hat a+1/2)}\hat S^{\dagger}| \psi(0)\rangle,\nonumber\
 \end{eqnarray}
and multiplying by $e^{-it \tilde{\omega}(\hat a^{\dagger}\hat a+1/2)}e^{it \tilde{\omega}(\hat a^{\dagger}\hat a+1/2)}$, we get
 \begin{equation}\label{1140}
| \psi(t)\rangle = e^{-it \tilde{\omega}(\hat a^{\dagger}\hat a+1/2)}e^{it \tilde{\omega}(\hat a^{\dagger}\hat a+1/2)}\hat Se^{-it \tilde{\omega}(\hat a^{\dagger}\hat a+1/2)}\hat S^{\dagger}| \psi(0)\rangle.
\end{equation}
We define now the time dependent squeezing operator
\begin{eqnarray}\label{1150}
\hat S_{\tilde{\omega}}(\lambda,t)&=&e^{it \tilde{\omega}(\hat a^{\dagger}\hat a+1/2)}\hat Se^{-it \tilde{\omega}(\hat a^{\dagger}\hat a+1/2)},\nonumber\\
&=&e^{it \tilde{\omega}(\hat a^{\dagger}\hat a+1/2)}e^{\frac{1}{8}\ln(1+\lambda) (\hat a^2-(\hat a^{\dagger})^2)}e^{-it \tilde{\omega}(\hat a^{\dagger}\hat a+1/2)},\nonumber\\
&=&\exp\Bigg[\frac{1}{8}\ln (1+\lambda) \big(e^{it \tilde{\omega}(\hat a^{\dagger}\hat a+1/2)}\hat a^2e^{-it \tilde{\omega}(\hat a^{\dagger}\hat a+1/2)}\nonumber\\
&-&e^{it \tilde{\omega}(\hat a^{\dagger}\hat a+1/2)}(\hat a^{\dagger})^2e^{-it \tilde{\omega}(\hat a^{\dagger}\hat a+1/2)}\big)\Bigg],\nonumber\
\end{eqnarray}
and we have
\begin{equation}\label{1160}
\hat S_{\tilde{\omega}}(\lambda,t) = \exp \left\{\frac{\ln (1+\lambda)}{8} \left[e^{-2it \tilde{\omega}}\hat a^2-e^{2it \tilde{\omega}}(\hat a^{\dagger})^2\right]\right\},
\end{equation}
taking (\ref{910}) the final form
\begin{equation}\label{1170}
| \psi(t)\rangle=e^{-it \tilde{\omega}(\hat a^{\dagger}\hat a+1/2)}\hat S_{\tilde{\omega}}(\lambda,t)\hat S^{\dagger}(\lambda)| \psi(0)\rangle.
\end{equation}
Up to  now we have an exact result. In order to compare with the approximation found with our method, we expand the operators in the above expression in Taylor series in terms of the "small" parameter $\lambda$.
First, we have to first order in $\lambda$,
\begin{equation}\label{1180}
e^{-it \tilde{\omega}(\hat a^{\dagger}\hat a+1/2)}= e^{-i t \frac{\tilde{\omega}}{\omega}\hat H_0}  \approx 1+(-it)\frac{\lambda}{2}  \hat H_0 .
\end{equation}
Second, we write
\begin{equation}\label{1190}
\hat S^{\dagger}(\lambda)=\sum_{n=0}^{\infty}\frac{\hat S^{{\dagger}(n)}(0)}{n!}\lambda^n;
\end{equation}
we keep first order terms, so
\begin{equation}\label{1200}
\hat S^{\dagger}(\lambda) \approx\hat S^{{\dagger}(0)}(0)+\lambda \hat S^{{\dagger}(1)}(0)
\end{equation}
and because
\begin{equation}\label{1210}
\hat S^{\dagger (0)}(0)=1,
\end{equation}
and
\begin{equation}\label{1220}
\hat S^{\dagger(1)} (0) =  \left[\frac{\partial S^{\dagger}(\lambda)}{\partial \lambda} \right]_{\lambda=0}    =\frac{((\hat a^{\dagger})^2-\hat a^2 )}{8},
\end{equation}
so to first order in $\lambda$
\begin{equation}\label{1230}
\hat S^{\dagger} (\lambda) \approx 1 + \lambda  \frac{((\hat a^{\dagger})^2-\hat a^2 )}{8}.
\end{equation}
Third, we write
\begin{equation}\label{1240}
\hat S_{\tilde{\omega}}(\lambda,t)=\sum_{n=0}^{\infty}\frac{\hat S^{ (n)}_{\tilde{\omega}}(0)}{n!}\lambda^n
\end{equation}
and it can be shown, that
\begin{equation}\label{1250}
\hat S^{(0)}_{\tilde{\omega}}(0)=1
\end{equation}
and that
\begin{equation}\label{1260}
\hat S^{ (1)}_{\tilde{\omega}}(0)=\left[\frac{\partial \hat S_{\tilde{\omega}}}{\partial \lambda}\right]_{\lambda=0}=
\frac{1}{8}[\hat a^2e^{-2it \omega}-(\hat a^{\dagger})^2e^{2it \omega}],
\end{equation}
so to first order in $\lambda$
\begin{equation}\label{1270}
\hat S_{\tilde{\omega}}(\lambda,t) \approx 1 + \lambda  \frac{1}{8}[\hat a^2e^{-2it \omega}-(\hat a^{\dagger})^2e^{2it \omega}].
\end{equation}
Finally, we arrive to the expression to order one in $\lambda$
\begin{equation}\label{1280}
| \psi(t)\rangle \approx e^{-it\hat H_0}\Bigg(1+\frac{\lambda}{8} \left( (\hat a^{\dagger})^2(1-e^{2i\omega t})+\hat a^2(e^{-2i\omega t}-1)\right)
-\lambda\frac{it\omega}{2}\bigg( \hat n+\frac{1}{2}\bigg)\Bigg)|\psi(0)\rangle,
\end{equation}
that when $|n^{(0)}\rangle$ is taken as initial condition, formula (\ref{710}) is recovered, that is the one obtained from the method introduced here.\\
The development to second order is long and bothersome and we do not present it here, but we recover the expressions found with our method in the previous section.

\section{The Dyson series in the matrix method}
It is well known that in terms of the Dyson series~  \cite{12}, the wavefunctions of the perturbed problem are written as
\begin{equation}\label{5010}
|\psi(t)\rangle=e^{-i\hat H_0 t}\hat T \left\{\exp\left[-i\lambda\int_{0}^{t}dt_1\hat V(t_1)\right]\right\}|\psi(0)\rangle.
\end{equation}
where $\hat T$ is the time order operator~  \cite{13}; i.e, if we have the time dependent operators $\hat A(t)$ and $\hat B(t)$ then
\begin{equation}\label{5020}
\hat T[\hat A(t_1) \hat B(t_2)]=\left\lbrace
  \begin{array}{ccc}
     \hat B(t_2)\hat A(t_1) \mbox{  if  }t_2>t_1, \\
     \hat A(t_1)\hat B(t_2) \mbox{  if  }t_1>t_2.
  \end{array}\right.
\end{equation}
On the other hand, from equation (\ref{560}) we can write
\begin{equation}\label{5030}
|\psi(t)\rangle= \left(e^{-i\hat{H}_0 t} +\sum_{n=1}^{\infty}\lambda^{n} \left( e^{-iMt} \right) _{(1,{n+1})}\right)|\psi(0)\rangle;
\end{equation}
so comparing equation (\ref{5010}) with equation (\ref{5030}), we derive the formula
\begin{equation}\label{5040}
\hat T\left\{\exp\left[-i\lambda\int_0^{t}dt_1 \hat V(t_1)\right]\right\}=e^{i\hat H_0 t}\sum_{n=0}^{\infty}\lambda^{n} \left( e^{-iMt} \right) _{(1,{n+1})}.
\end{equation}
which offers a matrix expansion for the Dyson operator and that links our matrix method with the Dyson series.

\section{Conclusions}
In this work, we have developed a new technique to find approximate solutions to the Schr\"odinger equation. We used the formal solution of the time dependent Schr\"odinger 
equation (\ref{130}). The key ingredient  is the introduction of the matrix $M$ defined in (\ref{580})  that allows us the transformation of the 
Taylor series for the wave function, in terms of products of the operators $\hat{H_0}$ and $\hat{V}$, in a series of powers of the matrix $M$, that is easier to handle.

The method allowed us to express the terms of the perturbation series in the form of integrals that depend on time that are restricted to the interval $[0,t]$, 
as appears in (\ref{310}) and (\ref{510}). An interesting property of these equations is that in the form that are presented they do not distinguish if the 
Hamiltonian $\hat{H_0}$ is degenerate or not, for what the equations that we provide for the corrections are very general expressions.

To test our method versus the standard Rayleigh-Schr\"odinger theory, we solved the corrections to the first and second order of quadratic potential as perturbation 
on the harmonic oscillator, and we found that the solution not only contains the results of the stationary case, but we find other time dependent terms that help us 
to study the temporal evolution of the corrections. It is important to remark that the terms as (\ref{037}), (\ref{540}), (\ref{541}) among others could not be found 
in the conventional treatment of the perturbation theory.

Finally, we have given  an alternative 
expression for the Dyson series in a matrix form.

\end{document}